\documentclass[11pt]{article}
\usepackage{fa2025}
\usepackage{amsmath}
\usepackage{cite}
\usepackage{url}
\usepackage{graphicx}
\usepackage{color}
\usepackage{siunitx}
\usepackage[utf8]{inputenc}
\usepackage{cite}
\usepackage{amsmath,amssymb,amsfonts}
\usepackage{algorithmic}
\usepackage{graphicx,color}
\usepackage{textcomp}
\usepackage{xcolor}

\usepackage{algorithm,algorithmic}
\usepackage[nolist]{acronym}
\usepackage{float}
\usepackage{amssymb}
\usepackage{etoolbox}
\usepackage{glossaries}
\usepackage{mathtools}
\usepackage{subcaption}
\usepackage{wrapfig}

\expandafter\def\expandafter\normalsize\expandafter{%
    \normalsize%
    \setlength\abovedisplayskip{0pt}%
    \setlength\belowdisplayskip{8pt}%
    \setlength\abovedisplayshortskip{-7pt}%
    \setlength\belowdisplayshortskip{3pt}%
}
\def\BibTeX{{\rm B\kern-.05em{\sc i\kern-.025em b}\kern-.08em
    T\kern-.1667em\lower.7ex\hbox{E}\kern-.125emX}}
\AtBeginDocument{\definecolor{tmlcncolor}{cmyk}{0.93,0.59,0.15,0.02}\definecolor{NavyBlue}{RGB}{0,86,125}}
\begin{acronym}
\acro{ATF}{Acoustic Transfer Function}
\acro{AED}{Auto Encoder-Decoder}
\acro{BCCTN}{Binaural Complex Convolutional Transformer Network}
\acro{BiTasNet}{Binaural TasNet}
\acro{BMWF}{Binaural MWF}
\acro{BRIRs}{Binaural Room Impulse Responses}
\acro{BSOBM}{Binaural STOI-Optimal Masking}
\acro{BTE}{behind-the-ear}
\acro{CED}{Convolutional Encoder-Decoder}
\acro{CNN}{Convolutional Neural Network}
\acro{CRM}{Complex Ratio Mask}
\acro{CRN}{Convolutional Recurrrent Network}
\acro{CASA}{Computational Auditory Scene Analysis}
\acro{DFT}{Discrete Fourier Transform}
\acro{DNN}{Deep Neural Network}
\acro{DOA}{Direction of Arrival}
\acro{ERB}{Equivalent Rectangular Bandwidth}
\acro{EVD}{Eigenvalue Decomposition}
\acro{FAL}{Frequency Attention Layer}
\acro{FTB}{Frequency Transformation Block}
\acro{FTM}{Frequency Transformation Matrix}
\acro{fwSegSNR}{Frequency-weighted Segmental SNR}
\acro{FCIM}{Fixed Cylindrical Isotropic MVDR}
\acro{GCC-PHAT}{Generalized Cross-Correlation with Phase Transform\acroextra{ method of estimating \acs{TDoA}}}
\acro{GRU}{Gated Recurrent Unit}
\acro{GOMVDR}{Guided OMVDR}
\acro{GFCIM}{Guided FCIM}
\acro{GiN}{Group in Noise}
\acro{GMVDR}{Guided MVDR}
\acro{GCC}{Generalized Cross-Correlation}
\acro{GMM}{Gaussian Mixture Model}
\acro{HATS}{Head and Torso Simulator}
\acro{HRIRs}{Head Related Impulse Response}
\acro{HSWOBM}{High-resolution Stochastic WSTOI-optimal Binary Mask}
\acro{IBM}{Ideal Binary Mask}
\acro{IDFT}{Inverse Discrete Fourier Transform}
\acro{ILD}{Interaural Level Difference}
\acro{IPD}{Interaural Phase Difference}
\acro{IRM}{Ideal Ratio Mask}
\acro{ISTFT}{Inverse STFT}
\acro{ITD}{Interaural Time Differences}
\acro{ISPD}{Interaural Signal Phase Difference}
\acro{iSNR}{input SNR}
\acro{LSA}{Log Spectral Amplitude}
\acro{LSTM}{Long Short-Term Memory}
\acro{MBSTOI}{Modified Binaural STOI}
\acro{MIMO}{Multiple Input Multiple Output}
\acro{MLP}{Multi-Layer Perceptrons}
\acro{MSC}{Magnitude Squared Coherence}
\acro{MVDR}{Minimum Variance Distortionless Response}
\acro{MWF}{Multichannel Wiener Filter}
\acro{MBCCTN}{Multichannel Binaural Complex Convolutional Transformer Network}
\acro{NCM}{Noise Covariance Matrix}
\acro{NLP}{Natural Language Processing}
\acro{NPM}{Normalized Projection Misalignment}
\acro{NH}{Normal Hearing}
\acro{OM-LSA}{Optimally-modified Log Spectral Amplitude}
\acro{OMVDR}{Oracle MVDR}
\acro{PESQ}{Preceptual Evalution of Speech Quality}
\acro{PReLU}{Parametric Rectified Linear Unit}
\acro{PSD}{Power Spectral Density}
\acro{ReLU}{Rectified Linear Unit}
\acro{RIR}{Room Impulse Responses}
\acro{RNN}{Recurrent Neural Network}
\acro{RTF}{Relative Transfer Function}
\acro{SI-SNR}{Scale-Invariant SNR}
\acro{SegSNR}{Segmental SNR}
\acro{SNR}{Signal-to-Noise Ratio}
\acro{SOBM}{STOI-optimal Binary Mask}
\acro{SPP}{Speech Presence Probability}
\acro{SSN}{Speech Shaped Noise}
\acro{STFT}{Short Time Fourier Transform}
\acro{STOI}{Short-Time Objective Intelligibility}
\acro{SRP}{Steered Response Power}
\acro{SRP-PHAT}{Steered Response Power with Phase Transform}
\acro{TF}{Time-Frequency}
\acro{TNN}{Transformer Neural Networks}
\acro{VAD}{Voice Activity Detector}
\acro{VSSNR}{Voiced-Speech-plus-Noise to Noise Ratio}
\acro{WGN}{White Gaussian Noise}
\acro{WSTOI}{Weighted STOI}

\end{acronym}
\title{Binaural Localization Model for Speech in Noise}
\multauthor
{Vikas Tokala$^{1*}$ \hspace{1cm} Eric Grinstein$^1$ \hspace{1cm} Rory Brooks$^1$} { \bfseries{Mike Brookes$^1$ \hspace{1cm} Simon Doclo$^2$ \hspace{1cm} Jesper Jensen$^{3,4}$ \hspace{1cm} Patrick A. Naylor$^1$}\\
  $^1$ Dept. of Electrical and Electronic Engineering, Imperial College London, UK\\
$^2$ Dept. of Medical Physics and Acoustics, Carl von Ossietzky Universität Oldenburg, Germany\\
$^3$ Demant A/S, Smørum, Denmark \\
$^4$ Dept. of Electronic Systems, Aalborg University, Denmark \\
\correspondingauthor{v.tokala@imperial.ac.uk}{Vikas Tokala et al.}
}

\begin{document}
% ---------------
% \name{Vikas Tokala$^{1}$, Eric Grinstein$^{1}$, Rory Brooks$^{1}$,
%       Mike Brookes$^{1}$, 
%       Simon Doclo$^{2}$,\\
%     Jesper Jensen${^{3,4}$,
%       Patrick A. Naylor}${^1} $\sthanks{This work was supported by funding from the European Union’s Horizon 2020 research and innovation programme under the Marie Skłodowska-Curie grant agreement No 956369 and the UK Engineering and Physical Sciences Research Council [grant number EP/S035842/1].}}
% \address{$^{1}$Dept. of Electrical and Electronic Engineering, Imperial College London, UK\\ $^{2}$ Dept. of Medical Physics and Acoustics, Carl von Ossietzky Universität Oldenburg, Germany.\\
%         $^{3}$  Demant A/S, Smørum, Denmark.\\
%         $^{4}$ Dept. of Electronic Systems, Aalborg University, Denmark}
%
% For example:
% ------------
%\address{School\\
%	Department\\
%	Address}
%
% Two addresses (uncomment and modify for two-address case).
% ----------------------------------------------------------
%\twoauthors
%  {A. Author-one, B. Author-two\sthanks{Thanks to XYZ agency for funding.}}
%	{School A-B\\
%	Department A-B\\
%	Address A-B}
%  {C. Author-three, D. Author-four\sthanks{The fourth author performed the work
%	while at ...}}
%	{School C-D\\
%	Department C-D\\
%	Address C-D}
%

% \ninept
%
\maketitle
\begin{abstract}
Binaural acoustic source localization is important to human listeners for spatial awareness, communication and safety. In this paper, an end-to-end binaural localization model for speech in noise is presented. A lightweight convolutional recurrent network that localizes sound in the frontal azimuthal plane for noisy reverberant binaural signals is introduced. The model incorporates additive internal ear noise to represent the frequency-dependent hearing threshold of a typical listener. The localization performance of the model is compared with the steered response power algorithm, and the use of the model as a measure of interaural cue preservation for binaural speech enhancement methods is studied. A listening test was performed to compare the performance of the model with human localization of speech in noisy conditions. 
\end{abstract}
\section{Introduction}
\label{sec:intro}
Binaural localization has garnered significant attention in the field of \ac{CASA}, which is influenced by principles underlying the perceptual organization of sound by human listeners. The two primary cues for sound localization are the \ac{ITD}, also known as the time difference of arrival, and the \ac{ILD}, which arises due to the influence of the head, torso, and outer ear. Differences between localization methods often stem from varying assumptions about environmental factors such as sound propagation, background noise, and microphone configuration. Localizing sound sources using binaural input in noise and reverberation is a challenging problem with important applications in hearing aids, spatial sound reproduction, and mobile robotics. 

It is well established that the noise and reverberation in typical listening environments can mask signals and negatively affect both binaural and monaural spectral cues, leading to reduced sound localization accuracy and speech comprehension even for individuals with normal hearing\cite{Good1996a,Lorenzi1999,Folkerts2023}. Research has shown that localization accuracy declines as the \ac{SNR} decreases. For instance, \cite{Good1996a} studied three normal hearing listeners who were asked to localize broadband click trains in an anechoic chamber under one quiet and nine noisy conditions with \ac{SNR}s ranging from -13 to +14 dB. Their findings revealed that localization accuracy was poorest in the lateral horizontal plane and began to deteriorate at \ac{SNR}s below +8~dB. Similarly, \cite{Lorenzi1999} investigated the effect of \ac{SNR} on localization ability in normal hearing listeners, finding that typical environments characterized by both noise and reverberation can further degrade localization cues and impair performance. In \cite{Kopco2010}, it is suggested that the combined effects of noise and reverberation could further reduce localization accuracy.
\begin{figure*}[!h]
    \centering
    \includegraphics[width=\textwidth]{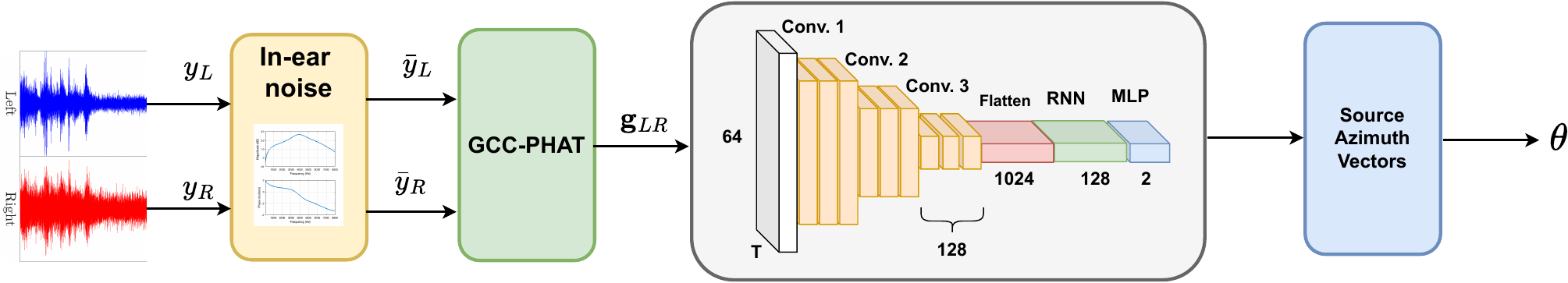}
    \caption{Block diagram of the model architecture.}
    \label{fig:blockDiag}
\end{figure*}
\begin{figure}[!h]
    \centering
    \includegraphics[width=0.4\textwidth]{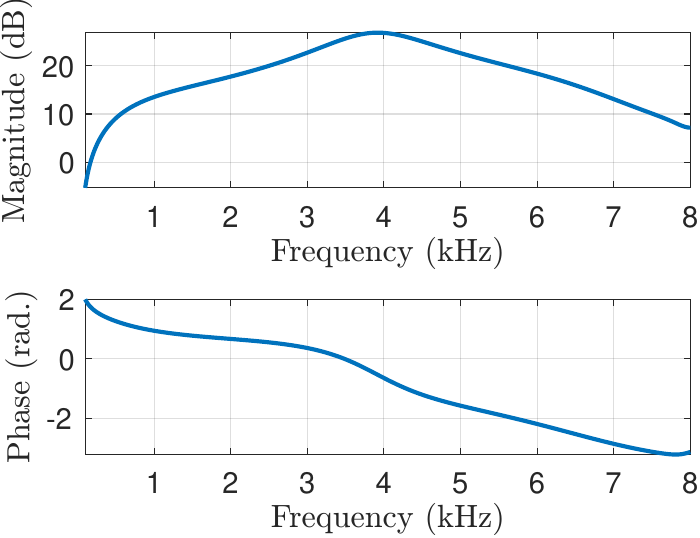}
    \caption{Magnitude and phase response of filter used to simulate the listener's hearing threshold.}
    \label{fig:inEarResp}
    % \vspace{-0.7cm}
\end{figure}
A well-known method for localisation using \ac{ITD} estimation is the \ac{GCC-PHAT} approach, which assumes ideal single-path propagation. Although \ac{GCC} and similar methods can be applied to any setup with two or more microphones, some recent research has focused on localization models specifically designed for binaural systems \cite{May2011, Ma2018}. Recent efforts have integrated azimuth-dependent models of \ac{ITD} and \ac{ILD}, demonstrating that jointly considering both cues enhances azimuth estimation compared to using \ac{ITD} alone\cite{Woodruff2012a,Ma2018,May2011}. However, these models often require prior training or calibration with the binaural input due to the significant variability in the frequency-dependent patterns of \ac{ITD}s and \ac{ILD}s across individuals, which can lead to performance degradation in different binaural setups. Methods also differ in how they integrate interaural information across time and frequency, with these variations largely reflecting different assumptions about source activity and interaction. In \cite{May2011}, authors proposed a framework that determines the likelihood of each source location based on a \ac{GMM} classifier, which learns the azimuth-dependent distribution of \ac{ITD}s and \ac{ILD}s from joint analysis of both binaural cues. However, many binaural localization methods have focused on scenarios with minimal reverberation or background noise. One approach to improving localization in more complex environments involves using model-based information about the spectral characteristics of sound sources in the acoustic scene to selectively weight binaural cues. This involves estimating models for both target and background sources during a training stage, using spectral features derived from isolated source signals \cite{Ma2018}. In \cite{Vecchiotti2019}, an end-to-end binaural localization algorithm that estimates the azimuth using \ac{CNN}s to extract features from the binaural signal was introduced. 

Human auditory cognition includes complex neurological processes for localization. Although \ac{ILD}s and \ac{ITD}s are widely accepted to be the primary interaural cues that influence human sound source localization \cite{Good1996a}, there is no standardized way to characterise them. Precedence effect, spectral cues, head movement and other psychoacoustical processes affect sound localization in humans. There is no universally accepted method of measuring the correlation between human sound localization and the frequency-varying interaural cues. In \cite{Tokala2023,Tokala2024}, to demonstrate the preservation of spatial cues, the error in interaural cues of the enhanced speech was computed using an \ac{IBM} that selects the speech-active regions in the signal.

A relevant approach to measuring the accuracy with which spatial information is preserved and the subsequent accuracy of localization of speech sources in noisy and enhanced speech signals would be to employ a model that predicts the localization of the speech-in-noise in a manner highly correlated to a human listener. This paper sets out to research methods of \ac{DOA} estimation that are not necessarily the best-performing but specifically follow the performance of the human listener in terms of binaural localization. The paper will focus on an end-to-end binaural localization model for speech in noisy and reverberant conditions, introducing a lightweight \ac{CRN} that utilizes input features based on \ac{GCC-PHAT}, which is a first step towards this goal. The model adds synthetic internal ear noise to an audio signal to simulate the effects of the frequency-dependent hearing threshold of a normal listener. The model is trained on binaural speech data to directly predict the source azimuth without limiting the localization to a predetermined azimuth-dependent distribution of interaural cues. The approach is evaluated using a listening test that was conducted using 15 normal hearing listeners, in which the participants were tasked to localize a target speaker in simulated noisy and reverberant conditions. 

% This paper is structured with the system description in Sec.~\ref{sec:sysDes} which describes the signal model, network architecture and loss function. Section~\ref{sec:experiments} describes the datasets, training and experimental setup followed by results and conclusion in Sec.~\ref{sec:results} and Sec.~\ref{sec:conclusion}. 

\section{System Description} \label{sec:sysDes}
\subsection{Signal model}
A binaural system is comprises a left and a right channel. The time-domain signal $y_{L}$ received by the left channel is modeled as
\begin{align}
    y_{L}(n) = s_{L}(n) + v_{L}(n),
\end{align}
where $s_{L}$ is the anechoic clean speech signal, $v_{L}$ is the noise and  $n$ is the discrete-time index. The in-ear noise added signal $\Tilde{{y}_L}$ is given by
\begin{align}
   \bar{y}_{L}(n) = h_e(n) \ast y_{L}(n) + e_{L}(n)
\end{align}
where $h_e(n)$ is the impulse response of the filter depicted in Fig.~\ref{fig:inEarResp} and  $e_{L}(n)$ is the white noise added to the filtered noisy signal. The right channel is described similarly with a $R$ subscript. The model adds fictitious internal ear noise to an audio signal to simulate the effects of the frequency-dependent hearing threshold of a normal listener, assuming that the input speech in the stronger channel is at the normal level defined in \cite{ANSI_S3_5_1997} to be 62.35~dB~SPL". The noise spectrum is taken from \cite{Pavlovic1987,ANSI_S3_5_1997} and, at a particular frequency, equals the pure-tone hearing threshold minus $10\log_{10}(C)$ where $C$ is the critical ratio. The critical ratio, $C$, is the power of a pure tone divided by the power spectral density of a white noise that masks it; this ratio is approximately independent of level. Hearing loss can also be taken into account here by modifying the filter that reduces the signal level by the hearing loss at each frequency. To avoid having to add very high noise levels at low and high frequencies, it instead filters the input signal by the inverse of the desired noise spectrum and then adds white noise with 0~dB power spectral density. Figure~\ref{fig:blockDiag} shows the block diagram of the proposed system. The raw time-domain signal is filtered with the in-ear frequency response shown in Fig.~\ref{fig:inEarResp}. The online implementation (\texttt{v\_earnoise.m} Matlab function) of the ear-noise filter can be found in \cite{Brookes1997}. The in-ear noise-added signal is then used as the input to the neural network, which determines the target azimuth in the frontal azimuthal plane.

% \begin{equation} \label{eq:gcc-phat}
%     {g}= \text{IDFT}\bigg(\frac{\bar{\mathbf{y}}_L}{\lvert \bar{\mathbf{y}}_L \rvert} \odot \frac{\bar{\mathbf{y}}^*_R}{\lvert \bar{\mathbf{y}}_R \rvert}\bigg),
% \end{equation}

\subsection{Localization network}

\subsubsection{Input Feature Set}
    The input feature of the proposed network consists of the \ac{GCC-PHAT} for the pair of microphone signal frames $(\mathbf{\bar{y}_L}, \mathbf{\bar{y}_R)}$, defined as
\begin{equation} \label{eq:gcc-phat}
    \mathbf{g}_{LR} = \text{IDFT}\bigg(\frac{\bar{\mathbf{Y}}_L}{\lvert \bar{\mathbf{Y}}_L \rvert} \odot \frac{\bar{\mathbf{Y}}^*_R}{\lvert \bar{\mathbf{Y}}_R \rvert}\bigg),
\end{equation}
% the \mbox{$L$-sized}
the \ac{IDFT} of the element-wise product of the normalized frequency-domain frames ${\mathbf{Y}_L}$ and $\mathbf{{Y}}_R$, where $\mathbf{\bar{{Y}}} = \text{DFT}(\bar{\mathbf{{y}}})$ and $\lvert {\mathbf{Y}} \rvert$ is the element-wise magnitude.

\subsubsection{Network architecture}
As shown in Fig.~\ref{fig:blockDiag}, the network is composed of a set of convolutional blocks, followed by an operation of flattening of the frequency and channel dimension. The resulting tensor is then used as input for a \ac{GRU} \ac{RNN}. Finally, a linear layer is applied to produce a 2-D output vector, $\mathbf{\hat{v}}$, representing the direction of the source's azimuth.  

\subsection{Loss function}
The proposed model is trained using a modification of the cosine similarity given by 
% \begin{align}
%     \cos(\mathbf{v}, \mathbf{\hat{{v}}}) = \frac{\mathbf{v} \cdot {\mathbf{\hat{v}}}}{|\mathbf{v}||{\mathbf{\hat{v}}}|}
% \end{align}
\begin{align} \label{eq:loss}
   \mathcal{L}({\mathbf{v}, \hat{\mathbf{v}}}) =  1 - \lVert \frac{\mathbf{v} \cdot {\mathbf{\hat{v}}}}{|\mathbf{v}||{\mathbf{\hat{v}}}|} \rVert
\end{align}
between the true and estimated directions $\mathbf{v}$ and $\hat{\mathbf{v}}$. The loss function \eqref{eq:loss} was designed so that the absolute value of the cosine similarity between the vectors is minimized, therefore not penalizing the effects caused by the front-back ambiguity, which are expected when employing only two microphones.
%\vspace{-0.5cm}

\section{Experiments}
\label{sec:experiments}
\subsection{Dataset}
To generate binaural speech data, monaural clean speech signals were obtained from the CSTR VCTK corpus \cite{Yamagishi2019} and spatialized using the measured \ac{BRIRs} from \cite{Francombe2017} for training. The VCTK corpus contains approximately 13 hours of speech data from 110 English speakers with various accents. These recordings were used to create 2~s speech utterances, which were spatialized to produce left and right ear channels. The resulting dataset comprised 20,000 speech utterances, which were divided into training (70\%), validation (15\%), and testing (15\%) sets. Diffuse isotropic speech-shaped noise was generated using uncorrelated noise sources uniformly distributed every $5^\circ$ in the azimuthal plane \cite{Moore2018b}, utilizing \ac{BRIRs} from \cite{Francombe2017} which were recorded in a listening room with a $T_{60}$ of $460$~ms. The binaural signals were generated with the target speech positioned at a random azimuth in the frontal plane ($-90 ^\circ$ to $+90^\circ$) with the source positioned at a distance of 100~cm. For the training process, isotropic noise was added so that the average in dB of $(SNR_L,~SNR_R)$, ranged between -25~dB and 25~dB. The evaluation set comprised speech signals spatialized with \ac{BRIRs} from \cite{Kayser2009} with random target azimuths and isotropic noise added at random \ac{SNR}s between -25~dB and 25~dB. The speaker was positioned at a $0^\circ$ elevation and at a distance of 3~m. This ensured that training and evaluation sets contained binaural signals generated using different \ac{BRIRs} to verify that the network generalised to different heads. 

\subsection{Training Setup}
The 2~s input signals were sampled at 16~kHz, and a window size of 512 was used to generate the signal frames with a 75\% overlap for a hop size of 25~ms. The parameters for the localization network are detailed in Fig.~\ref{fig:blockDiag}, which includes the tensor output shapes for each layer of the network. Convolutional layers employed a kernal size of (3, 3) throughout. Max pooling with a kernel size of 2 was applied to all convolutional layers except the last one. The \ac{PReLU} activation function was utilized in all layers of the network, except for the \ac{RNN} and  \ac{MLP} output layers, which used hyperbolic tangent ($\tanh$) activation, and the output layer, which employed sigmoidal activation. This architecture was taken from \cite{Grinstein2024a} and modified to work for binaural signals. The network has 850K parameters and is implemented using the PyTorch library, and the Adam optimizer was used for backpropagation. The network was trained for 80 epochs. The code for implementation is available online\footnote{\url{https://github.com/VikasTokala/BiL}}.

% \begin{align}
%    & R(\tau,\ell) = \sum_{k=0}^{K} \psi(k,\ell) G(k,\ell) e^{\frac{j2\pi f_s k \tau}{N}} & 
%    \end{align}
% where,
%    \begin{align}
%     &G(k,\ell) = S_L(k,\ell) S_R^*(k,\ell)& \\
%    & \psi(k,\ell) = \frac{1}{|G(k,\ell)|}&\\
%     &\tau_{max} = \argmax_{\tau}(R(\tau,\ell)) &\\
%     &\theta = \arcsin{\frac{c \cdot \tau_{max}}{d}}. &
% \end{align}

% \begin{itemize}
%     \item Details on the training and testing experiments
%     \item Mathematical info on Ear noise addition
%     \item Various experiments performed
    
% \end{itemize}

% \begin{figure}
%     \centering
%     \includegraphics[width=0.45\textwidth]{Figures/locError_perf.pdf}
%     \caption{Localization error.}
%     \label{fig:locErrorAn}
% \end{figure}
% \begin{figure}
%     \centering
%     \includegraphics[width=0.45\textwidth]{Figures/locError_methods.pdf}
%     \caption{Localization error for signals processed by different methods.}
%     \label{fig:locErrorRB}
% \end{figure}

% \begin{figure}
%     \centering
%     \includegraphics[width=0.45\textwidth]{Figures/locError_LisTest.pdf}
%     \caption{Localization error for listening tests compared with localization methods.}
%     \label{fig:locErrorLT}
% \end{figure}

\begin{figure*}[t]
     \centering
     \begin{subfigure}[b]{0.32\textwidth}
         \centering
         \includegraphics[width=0.95\textwidth]{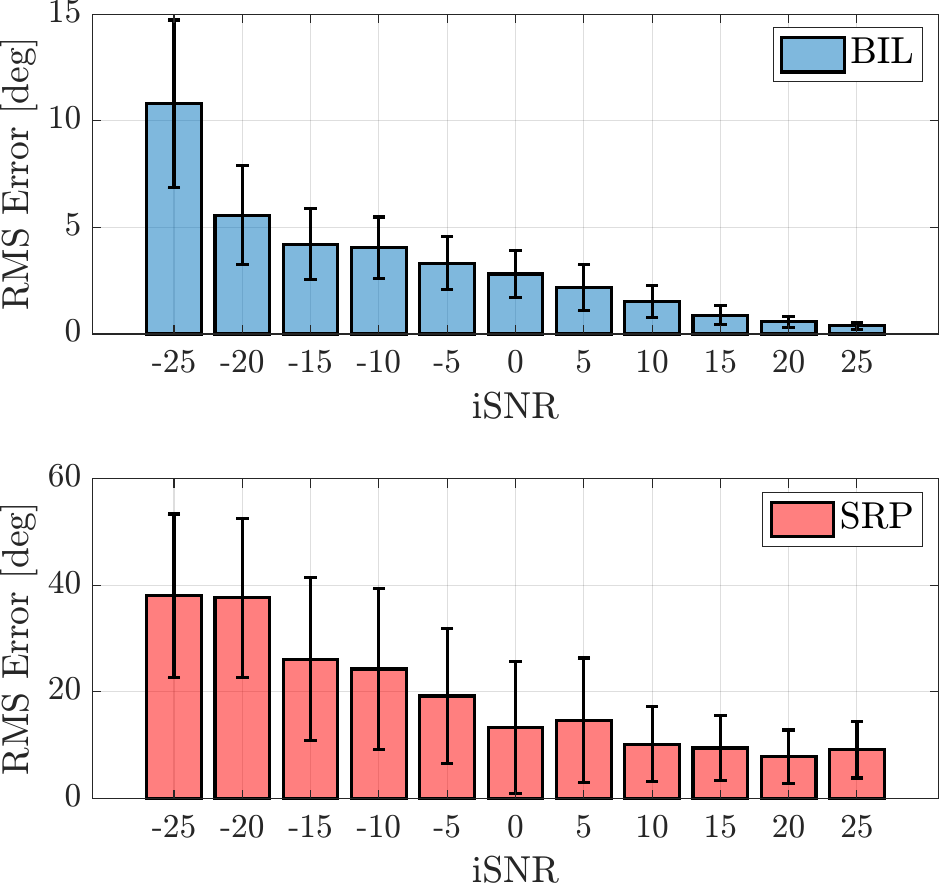} 
         \caption{}
         \label{fig:loc_error_perf}
     \end{subfigure}
     \hfill
     \begin{subfigure}[b]{0.33\textwidth}
         \centering
         \includegraphics[width=0.95\textwidth]{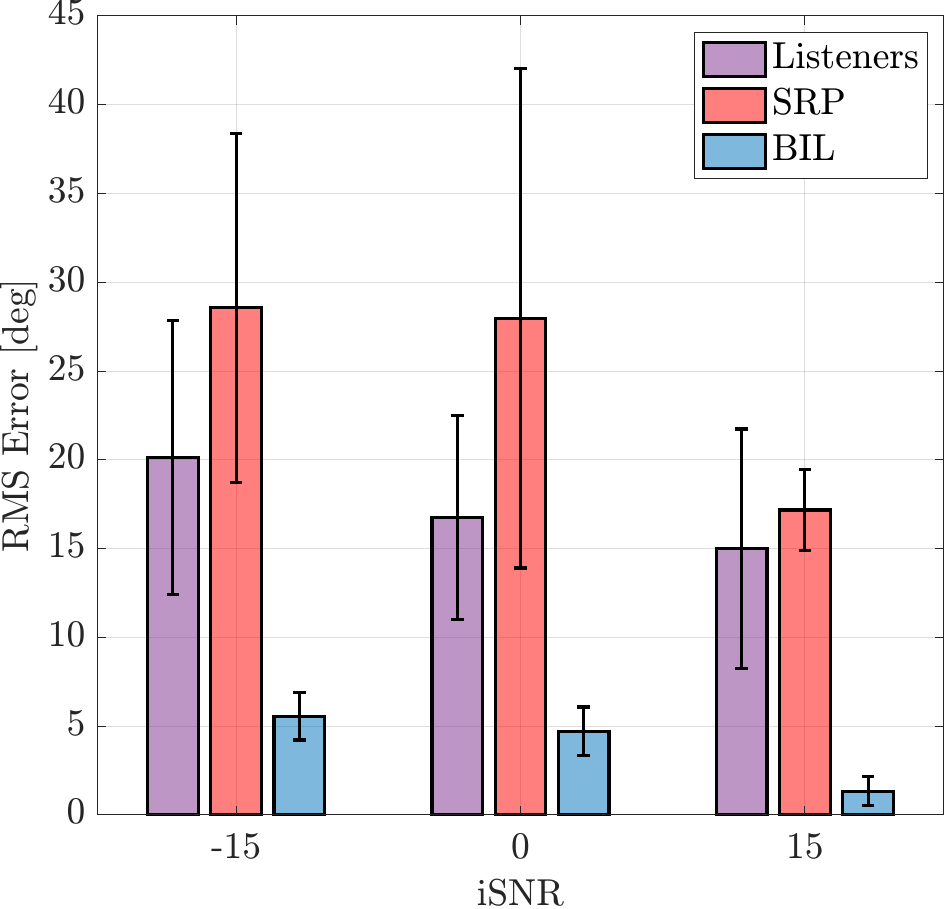}
         \caption{}
         \label{fig:loc_error_lisTest}
     \end{subfigure}
          \begin{subfigure}[b]{0.32\textwidth}
         \centering
         \includegraphics[width=0.95\textwidth]{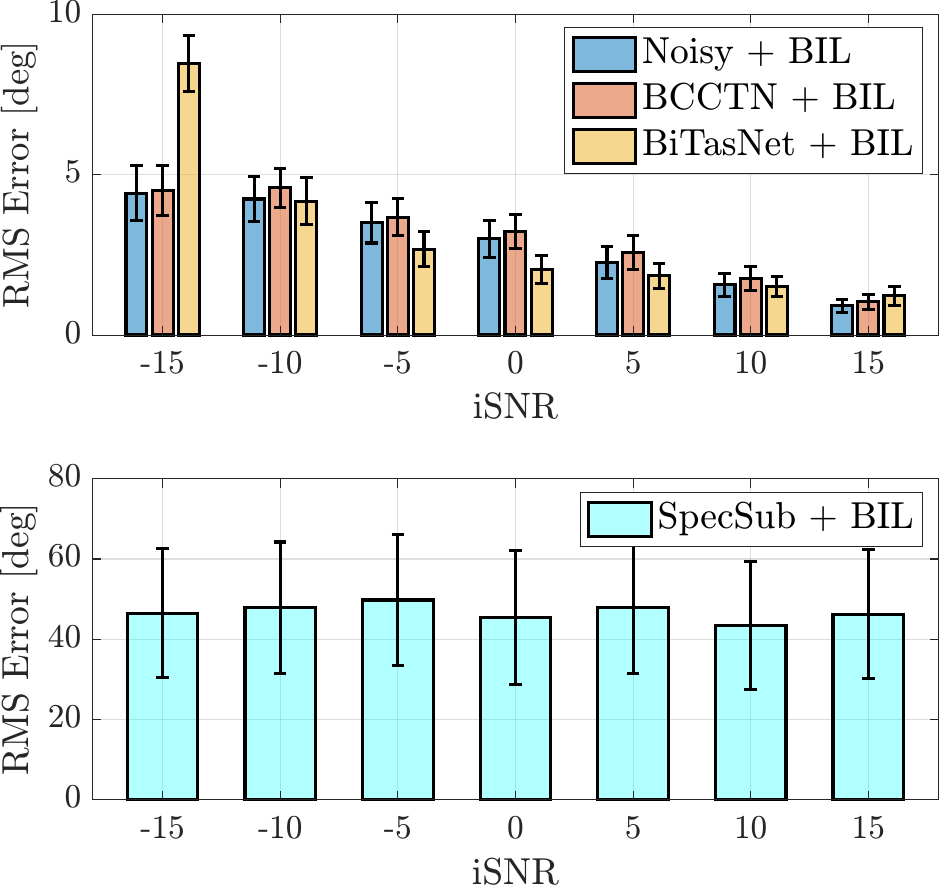}
         \caption{}
         \label{fig:loc_error_methods}
     \end{subfigure}
     \hfill
     %\vspace{-0.7cm}
        \caption{The plots show the localization error in noisy reverberant conditions (a) for the proposed method (BIL) and \acs{SRP}, (b) for listeners compared with the proposed method and \acs{SRP}  and, (c) for signals processed by different enhancement methods evaluated by BIL.}
        \label{fig:noises}
      % \vspace{-0.3cm}
\end{figure*}

\subsection{Listening Tests} \label{sec:lisTest}
In the listening tests, 15 participants with normal hearing were tasked with localizing a target speaker within the frontal azimuthal plane. Using Beyerdynamic DT1990 Pro open-back headphones, the audio signals were delivered in a soundproof booth through an RME Fireface UCX II audio interface. The participants were required to listen to the noisy speech utterances and select the perceived azimuth using a MATLAB-based GUI. The azimuths were quantized at $15^\circ$ intervals. Each participant listened to 36 speech utterances, which were evenly distributed across different \ac{SNR}s and randomly assigned azimuths in the frontal azimuth plane. Three conditions of \ac{iSNR} were used in the test: -15, 0 and +15~dB \ac{iSNR} corresponding to ``very noisy", ``noisy" and ``low noise" conditions, respectively.

\section{Results and Discussion}
\label{sec:results}
\begin{table}[]
\centering
\begin{tabular}{c|c}
\hline
\hline
Method         & Localization Error  \\ \hline \hline
SRP-PHAT       & $10.2^\circ$                           \\ \hline
WaveLoc-GTF \cite{Vecchiotti2019}  & $3.0^\circ$                          \\ \hline
WaveLoc-CONV \cite{Vecchiotti2019}& $2.3^\circ$                          \\ \hline
BIL            & $1.2^\circ$                          \\ \hline \hline
\end{tabular}
\caption{Localization error compared to WaveLoc \cite{Vecchiotti2019} methods.}
\label{tab:locError}
% \vspace{-0.8cm}
\end{table}

The model was evaluated using 275 speech utterances for each noisy input \ac{SNR} ranging from -25~dB to +25~dB in steps of 5~dB. The localization error for the proposed method, denoted as BIL, is shown in Fig.~\ref{fig:loc_error_perf} for different \ac{iSNR}s. The azimuth $\theta$ of the target speaker's \ac{DOA} in the frontal azimuth plane is then estimated using the \ac{SRP-PHAT} algorithm \cite{DiBiase2001a, Grinstein2024} and used for comparison. In extremely noisy conditions, such as -25~dB, the proposed method achieves a localization error of approximately $15^\circ$. Under similar \ac{iSNR} conditions, the localization error for \ac{SRP} is considerably higher, around $40^\circ$. As the \ac{iSNR} improves, the localization error for the proposed method decreases to below $5^\circ$, eventually reaching just under $1^\circ$ at 25~dB \ac{iSNR}. In contrast, the \ac{SRP} method maintains an error between $10^\circ$ and $20^\circ$ even at higher \ac{iSNR}s. The reduced performance of \ac{SRP} at higher \ac{iSNR}s can be attributed to reverberation, which causes multiple peaks in the correlation \cite{Grinstein2024a}. \ref{tab:locError} shows the comparison of localization error with the WaveLoc methods proposed in \cite{Vecchiotti2019}. These methods are also evaluated on \ac{BRIRs} from \cite{Francombe2017} without the addition of external noise, and the values shown are taken from \cite{Vecchiotti2019}. For similar conditions, the proposed method has lower error and outperforms both versions of the WaveLoc methods. 

Figure~\ref{fig:loc_error_lisTest} shows the localization error of human listeners compared with the proposed method and \ac{SRP} for the three conditions of noisy signals as described in Sec.~\ref{sec:lisTest}. The proposed method has a significantly lower localization error for all the \ac{iSNR} conditions. Listeners had an average error of $20^\circ$ in the very noisy condition of -15~dB and an average error of $15^\circ$ in the low noise condition of 15~dB, given that there was no head movement to assist them.  \ac{SRP}-based localization had the highest localization error and standard deviation for the test samples. Previous studies have shown that human localization of speech and tones can have a localization error of up to $40^\circ$ when noise and reverberation are present \cite{Good1996a, Lorenzi1999, Folkerts2023}. If the signals processed by enhancement methods produce a low localization error with the proposed method, it is very likely that the interaural cues of the signal are preserved, and human listeners will still localize the target speaker in the same azimuth as the original noisy signal.

Figure~\ref{fig:loc_error_methods} demonstrates how the proposed method can be used to assess the performance of binaural speech enhancement methods in preserving the interaural cues and the spatial information of the target speaker. While there are well-known objective measures to evaluate noise reduction, speech intelligibility and quality, there are no standardised measures to assess the preservation of binaural cues after they are processed by enhancement algorithms. The upper plot in Fig.~\ref{fig:loc_error_methods} shows the localization error for noisy signals at the \ac{iSNR}s from -15~dB to 15~dB and the signals processed by \ac{BCCTN} \cite{Tokala2023} and \ac{BiTasNet} \cite{Han2020} at the same \ac{iSNR}s.  The binaural enhancement algorithms are designed to preserve the interaural cues in the noisy signal while enhancement, and they show a low localization error. At -15~dB, the \ac{BiTasNet} shows a higher error compared to the noisy input signal, which indicates disruption in the interaural cues, and this is expected as the method was not designed to perform enhancement at -15~dB. As the \ac{iSNR} improves, all the binaural enhancement methods show localization error under $5^\circ$, which signifies the preservation of interaural cues. From Fig~\ref{fig:loc_error_perf} - Fig.~\ref{fig:loc_error_methods}, it is evident that the proposed model has a monotonic relationship to \ac{SNR}, i.e., the localization error decreases with increasing \ac{iSNR}. Furthermore, other studies, including \cite{Good1996a, Lorenzi1999, Folkerts2023}, show that human localization capability is monotonically proportional to \ac{SNR}. Hence, the proposed method has been seen to be, as desired, highly correlated with human binaural localization - a conclusion which is supported by the subjective listening tests conducted. The lower plot in Fig.~\ref{fig:loc_error_methods} shows the localization error obtained when the noisy signals are processed with bilateral spectral subtraction (SpecSub) \cite{Ephraim1985}, where no attempt is made at preserving binaural cues. The localization error is obtained around $45^\circ$ as the testset contains signals which have azimuths distributed randomly between $\pm90^\circ$. If the binaural enhancement methods are being used for purposes other than human listening, the addition of in-ear noise can be omitted before performing localization.

\section{Conclusion} \label{sec:conclusion}
This paper presented an end-to-end binaural localization model for speech in noisy and reverberant conditions. A \ac{CRN} network utilizing \ac{GCC-PHAT} features was introduced, and a listening test with 15 normal-hearing listeners showed that the model closely aligns with human perception, albeit with lower localization error. The model effectively evaluates the localization error of binaural speech enhancement algorithms, correlating with spatial information preservation and interaural cue retention. The key objective was to develop a \ac{DOA} estimation method that mirrors human binaural localization rather than purely optimizing accuracy. The proposed method demonstrated significantly lower localization errors across all \ac{iSNR} conditions. Listeners had average errors of $20^\circ$ at -15~dB and $15^\circ$ at 15~dB without head movement. \ac{SRP}-based localization showed the highest error and variability and as \ac{iSNR} improves, all binaural enhancement methods exhibit localization errors below $5^\circ$, confirming interaural cue preservation. The model’s localization error follows a monotonic relationship with \ac{SNR}, aligning with human performance trends.

\section{Acknowledgments}
This work was supported by funding from the European Union’s Horizon 2020 research and innovation programme under the Marie Skłodowska-Curie grant agreement No 956369 and the UK Engineering and Physical Sciences Research Council [grant number EP/S035842/1].
\bibliography{sapstrings,sapref}

\begin{thebibliography}{10}

\bibitem{Good1996a}
M.~D. Good and R.~H. Gilkey, ``Sound localization in noise: {{The}} effect of signal-to-noise ratio,'' {\em J Acoust Soc Am}, vol.~99, pp.~1108--1117, Feb. 1996.

\bibitem{Lorenzi1999}
C.~Lorenzi, S.~Gatehouse, and C.~Lever, ``Sound localization in noise in normal-hearing listeners,'' {\em J Acoust Soc Am}, vol.~105, pp.~1810--1820, Mar. 1999.

\bibitem{Folkerts2023}
M.~L. Folkerts, E.~M. Picou, and G.~C. Stecker, ``Spectral weighting functions for localization of complex sound. {{II}}. {{The}} effect of competing noise,'' {\em J Acoust Soc Am}, vol.~154, pp.~494--501, July 2023.

\bibitem{Kopco2010}
N.~Kop{\v c}o, V.~Best, and S.~Carlile, ``Speech localization in a multitalker mixture,'' {\em J Acoust Soc Am}, vol.~127, pp.~1450--1457, Mar. 2010.

\bibitem{May2011}
T.~May, S.~{van de Par}, and A.~Kohlrausch, ``A {{Probabilistic Model}} for {{Robust Localization Based}} on a {{Binaural Auditory Front-End}},'' {\em {IEEE/ACM} Trans. Audio, Speech, Language Process.}, vol.~19, pp.~1--13, Jan. 2011.

\bibitem{Ma2018}
N.~Ma, J.~A. Gonzalez, and G.~J. Brown, ``Robust {{Binaural Localization}} of a {{Target Sound Source}} by {{Combining Spectral Source Models}} and {{Deep Neural Networks}},'' {\em {IEEE/ACM} Trans. Audio, Speech, Language Process.}, vol.~26, pp.~2122--2131, Nov. 2018.

\bibitem{Woodruff2012a}
J.~Woodruff and D.~Wang, ``Binaural {{Localization}} of {{Multiple Sources}} in {{Reverberant}} and {{Noisy Environments}},'' {\em {IEEE/ACM} Trans. Audio, Speech, Language Process.}, vol.~20, pp.~1503--1512, July 2012.

\bibitem{Vecchiotti2019}
P.~Vecchiotti, N.~Ma, S.~Squartini, and G.~J. Brown, ``End-to-end {{Binaural Sound Localisation}} from the {{Raw Waveform}},'' in {\em Proc. {IEEE} Int. Conf. on Acoust., Speech and Signal Process. ({ICASSP})}, pp.~451--455, May 2019.

\bibitem{Tokala2023}
V.~Tokala, E.~Grinstein, M.~Brookes, S.~Doclo, J.~Jensen, and P.~A. Naylor, ``Binaural {{Speech Enhancement}} using {{Deep Complex Convolutional Recurrent Networks}},'' in {\em Proc. Asilomar Conf. on Signals, Syst. \& Comput.}, (USA), 2023.

\bibitem{Tokala2024}
V.~Tokala, E.~Grinstein, M.~Brookes, S.~Doclo, J.~Jensen, and P.~A. Naylor, ``Binaural {{Speech Enhancement}} using {{Deep Complex Convolutional Transformer Networks}},'' in {\em Proc. {IEEE} Int. Conf. on Acoust., Speech and Signal Process. ({ICASSP})}, (Seoul, South Korea), 2024.

\bibitem{ANSI_S3_5_1997}
{ANSI}, ``Methods for the calculation of the speech intelligibility index,'' {{ANSI Standard}} S3.5-1997 (R2007), American National Standards Institute (ANSI), 1997.

\bibitem{Pavlovic1987}
C.~V. Pavlovic, ``Derivation of primary parameters and procedures for use in speech intelligibility predictions,'' {\em J Acoust Soc Am}, vol.~82, pp.~413--422, Aug. 1987.

\bibitem{Brookes1997}
D.~M. Brookes, ``{{VOICEBOX}}: {{A}} speech processing toolbox for {{MATLAB}},'' 1997.

\bibitem{Yamagishi2019}
J.~Yamagishi, C.~Veaux, and K.~MacDonald, ``{{CSTR VCTK Corpus}}: {{English}} multi-speaker corpus for {{CSTR}} voice cloning toolkit (version 0.92),'' {\em University of Edinburgh. The Centre for Speech Technology Research (CSTR)}, 2019.

\bibitem{Francombe2017}
J.~Francombe, ``{{IoSR Listening Room Multichannel BRIR Dataset}} - {{University}} of {{Surrey}},'' 2017.

\bibitem{Moore2018b}
A.~H. Moore, L.~Lightburn, W.~Xue, P.~A. Naylor, and M.~Brookes, ``Binaural mask-informed speech enhancement for hearing aids with head tracking,'' in {\em Proc. {{Int}}. {{Workshop}} on Acoust. {{Signal}} Enhancement ({{IWAENC}})}, (Tokyo, Japan), pp.~461--465, Sept. 2018.

\bibitem{Kayser2009}
H.~Kayser, S.~D. Ewert, J.~Anem{\"u}ller, T.~Rohdenburg, V.~Hohmann, and B.~Kollmeier, ``Database of multichannel in-ear and behind-the-{{Ear}} head-related and binaural room impulse responses,'' {\em EURASIP J. on Advances in Signal Process.}, vol.~2009, p.~298605, July 2009.

\bibitem{Grinstein2024a}
E.~Grinstein, C.~M. Hicks, T.~{van Waterschoot}, M.~Brookes, and P.~A. Naylor, ``The {{Neural-SRP Method}} for {{Universal Robust Multi-Source Tracking}},'' {\em IEEE Open Journal of Signal Processing}, vol.~5, pp.~19--28, 2024.

\bibitem{DiBiase2001a}
J.~H. DiBiase, H.~F. Silverman, and M.~S. Brandstein, ``Robust localization in reverberant rooms,'' in {\em Microphone {{Arrays}}} (M.~Brandstein and D.~Ward, eds.), Digital {{Signal Processing}}, pp.~157--180, Berlin Heidelberg: Spring\-er-Ver\-lag, 2001.

\bibitem{Grinstein2024}
E.~Grinstein, E.~Tengan, B.~{\c C}akmak, T.~Dietzen, L.~Nunes, T.~{van Waterschoot}, M.~Brookes, and P.~A. Naylor, ``Steered {{Response Power}} for {{Sound Source Localization}}: {{A Tutorial Review}},'' {\em {EURASIP} J. on Audio, Speech, and Music Process.}, vol.~submitted, May 2024.

\bibitem{Han2020}
C.~Han, Y.~Luo, and N.~Mesgarani, ``Real-{{Time Binaural Speech Separation}} with {{Preserved Spatial Cues}},'' in {\em Proc. {IEEE} Int. Conf. on Acoust., Speech and Signal Process. ({ICASSP})}, pp.~6404--6408, May 2020.

\bibitem{Ephraim1985}
Y.~Ephraim and D.~Malah, ``Speech enhancement using a minimum mean-square error log-spectral amplitude estimator,'' {\em IEEE Trans. Acoust., Speech, Signal Process.}, vol.~33, no.~2, pp.~443--445, 1985.

\end{thebibliography}

\end{document}